\documentclass{sigchi}

\toappear{Permission to make digital or hard copies of all or part of this work for personal or classroom use is granted without fee provided that copies are not made or distributed for profit or commercial advantage and that copies bear this notice and the full citation on the first page. Copyrights for components of this work owned by others than the author(s) must be honored. Abstracting with credit is permitted. To copy otherwise, or republish, to post on servers or to redistribute to lists, requires prior specific permission and/or a fee. Request permissions from Permissions@acm.org.\\
  {\emph{CSCW'17}}, February 25-March 01, 2017, Portland, OR, USA. \\
  Copyright is held by the owner/author(s). Publication rights licensed to ACM.\\
  ACM 978-1-4503-4335-0/17/03...\$15.00\\
  DOI: http://dx.doi.org/10.1145/2998181.2998354}

\usepackage{balance}  
\usepackage{graphics} 
\usepackage{txfonts}
\usepackage{times}    
\usepackage[pdftex]{hyperref}
\usepackage{color}
\usepackage{textcomp}
\usepackage{booktabs}
\usepackage{ccicons}
\usepackage{todonotes}
\usepackage{csquotes}
\usepackage{acmcopyright}

\makeatletter
\def\url@leostyle{%
  \@ifundefined{selectfont}{\def\UrlFont{\sf}}{\def\UrlFont{\small\bf\ttfamily}}}
\makeatother
\def\pprw{8.5in}
\def\pprh{11in}

\definecolor{linkColor}{RGB}{6,125,233}
\hypersetup{%
  pdftitle={Bots as Virtual Confederates: Design and Ethics},
  pdfauthor={Peter Krafft, Michael Macy, Alex Pentland},
  pdfkeywords={oline field experiments; bots; virtual confederates; ethics},
  bookmarksnumbered,
  pdfstartview={FitH},
  colorlinks,
  citecolor=black,
  filecolor=black,
  linkcolor=black,
  urlcolor=linkColor,
  breaklinks=true,
}

\begin{document}

\title{Bots as Virtual Confederates: Design and Ethics}

\numberofauthors{1}
\author{%
  \alignauthor{Peter M. Krafft$^*$, Michael Macy$^\dagger$, Alex ``Sandy'' Pentland$^*$\\
    \affaddr{$^*$Massachusetts Institute of Technology, $^\dagger$Cornell}\\
    \email{pkrafft@mit.edu, m.macy@cornell.edu, pentland@mit.edu}}\\
}


\maketitle

\begin{abstract}
  The use of bots as virtual confederates in online field experiments
  holds extreme promise as a new methodological tool in computational
  social science.  However, this potential tool comes with inherent
  ethical challenges.  Informed consent can be difficult to obtain in
  many cases, and the use of confederates necessarily implies the use
  of deception.  In this work we outline a design space for bots as
  virtual confederates, and we propose a set of guidelines for meeting
  the status quo for ethical experimentation.  We draw upon examples
  from prior work in the CSCW community and the broader social science
  literature for illustration.  While a handful of prior researchers
  have used bots in online experimentation, our work is meant to
  inspire future work in this area and raise awareness of the
  associated ethical issues.
\end{abstract}



\section{Introduction}

Randomized experiments provide the gold standard for evidence of causality in the behavioral sciences.  Yet many questions in social science are difficult to study experimentally. The
use of randomized trials to study human social behavior requires
situating human subjects in settings where the attributes and behavior
of co-present individuals are controlled.

Traditionally, social psychologists and sociologists have often addressed
this problem through the use of ``confederates'' who are trained by
the researcher to follow pre-assigned scripts. For example, Solomon
Asch, one of the pioneers in the use of confederates in social
psychology, tested conformity by exposing participants to
incorrect perceptual judgments of other group members, all of whom
were confederates \cite{asch1955opinions}. Stanley
Milgram, Asch's student, used confederates in a field experiment by having them stand on the sidewalk and direct
their gaze upwards towards a tall building. Passersby responded by
also looking up \cite{milgram1969note}.  Using human confederates to
manipulate social situations works well for certain laboratory
experiments and small field experiments, but is not scalable to
long-running or large-scale field experiments involving thousands of
participants, and the personalities and demographics of human
confederates remain difficult to control.


Bots as ``virtual confederates'' (or ``confedabots'') provide a
powerful solution that addresses the limitations of human
confederates.  
Bots are digital agents who act according to tailored algorithms, often
as peers, in online spaces such as online social networks or
chatrooms.  Bots are increasingly prevalent in commercial and hobbyist
applications, but are also beginning to make appearances in the
academic literature as tools for scientific experimentation.  For
instance, several groups of researchers have used bots to test which
attributes and activity patterns lead to greater numbers of followers
on Twitter and other online social networks
\cite{aiello2012people,messias2013you,freitas2015reverse}. Others have
proposed using Twitter bots to bridge polarized communities and pop
filter bubbles \cite{graham2016do}. Within the CSCW community,
researchers recently used bots to test which social media strategies
might be most effective at mobilizing volunteer activists
\cite{savage2016botivist}.
Still, the use of bots for scientific experimentation has
promise that potentially extends far beyond this handful of existing cases.


Bots allow for fine-grained control and high degrees of research
replicability and transparency (via the publication of bot source
code). Demographic attributes, personalities, and normative behaviors
can be randomly assigned to bots. Since bot behavior is completely
automated, bots can also be used for large-scale and longitudinal
social experimentation in which a large number of confederates must
engage with participants for a long period of time.


CSCW and the surrounding communities are likely to use confedabots in
the future due to the existing expertise in the necessary areas and
due to the need for this technique.  Prior work could have potentially
been strengthened by using bots for causal inference.  Researchers
have used observational analysis to study the predictors of following
behavior on Twitter \cite{hutto2013longitudinal}.  Bots could be used to
test the causality of these findings
\cite{messias2013you,freitas2015reverse}.  Other recent work has
examined why people make particular choices in Doodle polls
\cite{zou2015strategic}.  Similar studies could be conducted using
bots to manipulate the popularity of time slots in order to remove
confounds in the natural experiment the prior research analyzed.
Researchers are also interested in how people respond to requests for
help on social media.  A recent study identifying differences in how
people respond to these requests versus other types of social media
posts could have also been complemented by bots making these different
requests \cite{lampe2014help}.


Alongside these largely unexplored potential applications of bots as
virtual confederates, however, there are ethical challenges.  Although
not specifically related to bots as virtual confederates, major public
outcries against two online field experiments have already occurred
due to the violation of public expectations about researcher behavior.
When Facebook published the results of their notorious ``emotional
contagion'' experiment in 2014 \cite{kramer2014experimental}, a bustle
of negative attention ensued.  One common complaint was that Facebook
had not sought informed consent.  Another issue was the possibility of
harm stemming from the researchers' attempts to manipulate users'
emotions.  A second scandal occurred when Microsoft Research released
a Twitter bot called ``Tay'' that interacted with users on Twitter and
learned from those interactions.  Due to its learning algorithm and
the content it was exposed to, Tay ultimately began posting disturbing
and offensive content.

Our goals in this paper are to outline a potential design space for
future confedabot experiments, and to discuss ways to avoid 
ethical issues with these types of experiments.  The design space we
propose is grounded in the existing work that has used bots, but we
hope to also expose the substantially broader possibilities of this
technique.  A number of previous authors have discussed the ethics of
online experimentation, but none  have addressed the challenges
associated with bots and how to overcome them.  After outlining our design space for bots as virtual
confederates, we draw upon extensive prior exploration of the ethics
of related experimental techniques in order to explicate the major
ethical concerns with using bots as virtual confederates in online
field experiments.  We use precedent set by existing practices
and community norms in order to propose guidelines for how to design
ethical field experimentation involving confedabots.

\section{Background}

\subsection{Online Field Experiments}

Online field experiments consist of randomized trials conducted on
websites or other networked digital platforms such as Facebook, Twitter,
or LinkedIn. One of the first high-profile
massive online field experiments implemented a system on Facebook that
encouraged U.S. users to vote in a general election
\cite{bond201261}. The research found that showing users pictures of
their friends who had voted was far more effective in promoting voter
turnout than a message from Facebook reminding users to vote.  However, this experiment did not involve bots.

\subsection{Virtual Confederates}

Virtual confederates are artificial agents who act like human
confederates in an experimental context.  The concept of a virtual
confederate has been explored previously in laboratory settings
\cite{blascovich2002immersive}, but not in field experiment settings.
The use of virtual confederates in online field experiments presents
challenges not faced in the lab.


\subsection{Bots}

A bot is an
algorithm that automatically generates user content and interactions
in an online space.  Many online spaces are already populated by bots.
For example, at least 10\% of Twitter users are thought to be bots of
one kind or another.\footnote{Source: Quarterly Twitter SEC Form 10-Q,
  June 30, 2014. Commission File Number: 001-36164.}  Many of these
bots are used by companies or other organizations for direct
advertisement, search engine optimization, or other promotional
purposes.  Others of these bots are interesting, creative, and highly
valued by the community.  For example, @congressedits is a Twitter bot
that monitors wikipedia for edits made by IP addresses that can be
traced to members of the U.S. Congress.

\subsection{Internet Ethics}

The ethics of internet research continues to be an area of sustained
interest in the CSCW community
\cite{fisher2010terms,fiesler2015ethics,keegan2016actually} and beyond
\cite{kraut2004psychological,buchanan2012internet,salganik2017bit}.
As anticipated by the foundational literature in this area
\cite{markham2012ethical}, the techniques of internet research
continue to evolve, and we must continue the conversation around the
ethics of those techniques.  To the best of our knowledge, there have
been no investigations yet into the ethics of bots as virtual
confederates for online field experiments.


\section{Bots as Virtual Confederates}

In this section we provide a scheme for conceptualizing
confedabot experiments.  We outline four categories of
online field experiments that could be conducted using bots: varying
actions, varying attributes, varying algorithms, and creating
artificial social contexts.



\subsection{Intervening on Actions}

The simplest kind of experiment to do with a bot is to randomize
individual actions to identify the effects of those actions.  These
interventions can either be isolated, so that the only purpose of the
bot is to execute these actions, or these randomized interventions can
be embedded within the regular behavior of the bot.



\paragraph{Example}

A simple example of such a bot is one that implements a
rich-get-richer field experiment.  A number of researchers have
studied how incrementing the popularity of items in online systems can
lead to gains in future popularity (e.g.,
\cite{muchnik2013social,van2014field}).  A bot implementing such an
experiment would look for random pieces of content to upvote, for
example on a site like reddit.

Another example of this type of experiment is one where a subset of
the bot's actions form the experiment itself, and the rest of the
actions form an identity for the bot.  Aside from the fact that the
bots were not pretending to be humans, a good example of this type of
experiment is the prior work examining which volunteer recruitment
strategies are most useful for Twitter bots \cite{savage2016botivist}.
These researchers created bots who would follow a simple workflow to
attract volunteers, and the wording of their initial message to
potential volunteers was randomized across four types.






\subsection{Intervening on Attributes}

A second type of experiment is to intervene on an attribute of a bot
that doesn't actually affect the bot's behavior.  In this case, the
algorithm generating the bot's actions is a unit of control, and the
experiment is layered on top of the algorithm generating that content.



\paragraph{Example}

A prime example of intervening on attributes is a minimal version of the
type of bots used in prior work to reverse engineer the techniques used by popular Twitter
bots to gain followers \cite{freitas2015reverse}.  A minimal version of this Twitter bot would always
tweet random content (e.g., quotes from other users or content
generated from an n-gram language model), but some attribute of the
account, such as the presented gender of the bot, would be randomized.


\subsection{Intervening on Algorithms}

A third type of experiment is to intervene on the entire behavioral
profile of a bot.  In this case, the effect of a particular behavioral
profile as embodied in the bot's algorithm is tested.



\paragraph{Example}

An example of this sort of bot would be a different minimal version of
the type of bots used in prior work to reverse engineer popular
Twitter bots \cite{freitas2015reverse}.  This minimal Twitter bot
would again always tweet random content, but would do so at either a
high rate or a low rate.  This experiment tests the effect of activity
rate on following behavior.  Another example would be to vary the type
of content that the bot tweets about, or the personality of the bot.


\subsection{Artificial Contexts}

A final type of experiment consists of the creation of an artificial
context where multiple bots are used together in order to create a
rich social scenario.

\paragraph{Example}
One potential example would be to create a
forum thread for the purposes of an experiment, and populate it
initially with multiple bots who interact with each other in that
thread.


\section{Basic Ethical Issues}

In the United States, the Belmont Report is the canonical document
that lays out the basic guidelines for ethical experimentation in the
behavioral sciences.  The Belmont Report centers around three
principles: \textbf{respect for persons}, \textbf{beneficence}, and
\textbf{justice}.  Following these three principles ensures that the
personal autonomy of participants is not violated, that the benefits
of a study outweigh its risk, and that the benefits and risks are
distributed fairly among the participant population.  These principles
have been codified in policy within the United States in what is
called the ``Common Rule''.  In this section we outline how the principles
described in the Belmont Report, as codified in the Common Rule, come
to bear on the major ethical issues involved in using bots as virtual
confederates in online field experiments.  By and large, we are able
to focus our discussion around these foundational principles because
most modern ethical guidelines for experimentation closely mirror the
principles outlined in the Belmont Report.  Nonetheless, there is
substantial disciplinary variation in how conservatively these
principles are interpreted and in how they are enforced.
Where there is notable disciplinary variation as it concerns bots as
confederates, we will add discussion.









\subsection{Informed Consent}

The ethical principle of respect for persons implies that researchers
must obtain informed consent in human experimentation---experimenting
on participants without informed consent necessarily entails the
intention to violate those participants' personal autonomy.  Unfortunately informed consent can
be difficult or impossible to obtain in many online field experiments, especially
those conducted as a peer in an online system.  The content a bot creates is likely publicly available, and thus the researcher cannot
control who gets to see that content.  Even in settings
where the researcher can control information flow, asking users via direct messages for permission to
expose them to the experiment might be more
intrusive than conducting the experiment itself.

Perspectives differ on whether there can be exceptions to the
requirement that informed consent be obtained.  Recent work
investigating the ethical practices of CSCW researchers found that
22\% of researchers surveyed held the view that obtaining informed
consent is always necessary \cite{vitak2016beyond}.  At the same
time, waivers of informed consent are sought and obtained in field
experiments across the social sciences
\cite{levitt2009field,chen2015online}.  However, the omission of informed consent
was one of the key controversial issues discussed in the turmoil
surrounding Facebook's emotional contagion experiment.


Practically speaking, many confedabot experiments will only be
possible if the requirement for informed consent is waived.  The
Common Rule provides explicit guidelines on when such a waiver may be
permissible: when the risks of an experiment are extremely low and
obtaining informed consent is difficult.  Omitting informed consent in
an experiment that uses bots as virtual confederates must therefore
clearly pose minimal risk.  In developing our guidelines for
the viable use of confedabots, this consideration will be paramount.

\subsection{Deception}

Strictly interpreted, the principle of respect for persons prohibits
most forms of deception in behavioral experimentation,  unless
a participant has consented to being deceived.  Deception is
problematic because it is used with the expectation that participants
would behave differently were they not deceived.  Deception therefore
intentionally circumvents individual autonomy.  

Unfortunately, the use of confedabots is likely to entail some degree
of deception.  Predominantly, for the conclusions gleaned from
experiments using confedabots to be compelling, the researcher may
hope that people think the confedabots are actually human users.
Explicit deception is also a possibility.  Bots that spread
misinformation about facts or world events could be considered
unethical.  More subtle grey cases can also occur.  Even if a bot is
not explicitly spreading false information about the world, the
actions or words of a bot might still be able to viewed as explicitly
deceptive if the bot references its internal mental states.  In an
extreme example, a confedabot expressing love for a person appears
intuitively unethical.  In this case we do not believe the bot
actually has the internal feeling of love, and hence the bot is lying
about loving the person.  In less extreme cases, if a bot copies the
tweets of a human, and that human expresses certain internal feelings
or references particular personal life events, the bot would be lying
again.  Even in the banal case of a bot that simply upvotes random
content, that bot is being deceptive about its judgements of what
it likes since the bot has no internal preferences about what to
upvote.

Perspectives differ widely across fields on the ethics of using
deception in behavioral experimentation.  In the field of psychology,
deception is considered permissible, but the relative benefits of
deception must justify its use, and participants must eventually be
told that they were deceived \cite{american2002ethical}.  In
economics, the use of deception is generally forbidden.
Interestingly, the main reason that economists frown upon deception is
not directly out of ethical considerations for the participants, but
rather out of concern for the effect that the use of deception might
ultimately have on the validity of behavioral experiments.  One of the
main concerns is maintaining trust within the participant population
\cite{davis1993experimental}.  The concern is that experiments
involving deception may adversely affect how participants behave not
only in the deceitful experiments but also in those that do not
involve deception, because participants will always expect they are
being tricked.  In principle, this justification would seem to exclude
any form of deception.  In practice, however, there is a meaningful
difference between explicit deception versus deception by omission
\cite{krawczyk2013delineating}.  Surveys by economists of researchers
and participants show that explicit deception is thought of as being
less acceptable than deception by omission
\cite{krawczyk2013delineating,rousu2015deception}.  Furthermore,
deception by omission has been used in a number of relatively recent
influential field experiments in economics
\cite{bertrand2004emily,camerer1998can,riach2002field}.


There is documented disagreement among CSCW researchers on the use of
deception. Researchers in communications tend to feel more comfortable
with the use of deception than information scientists or computer
scientists \cite{vitak2016beyond}.  A conservative reading of the ACM
Code of Ethics, the canonical ethics document for researchers in
computer science, forbids the use of deception.  ACM charges their
members to ``be honest and trustworthy'' as a moral imperative.  The
explicit guidelines associated with this imperative are mostly
targeted at engineers who build products, apparently to be honest and
trustworthy towards their customers, but the principle itself could
easily be interpreted more broadly.

In terms of the views of the public at large, one case of early work
studying how bots can become popular found that some people on the
site being studied were unhappy when they discovered a bot had become
so influential \cite{aiello2012people}.  The bot was banned by site
administrators, users expressed discomfort at the unfamiliar account
frequently visiting their profiles, and users expressed concern of
privacy violation.

Given the complexities involved in these different forms of possible
deception, and the lack of a consensus for what types of deception are
permissible, we recommend the Association of Internet Researchers'
(AoIR) case-by-case approach of benefit-harm analysis
\cite{markham2012ethical} in analyzing what amount of deception is
permissible. In many cases, it may also be possible to announce the use of deception on the bot profile at the conclusion of the experiment to satisfy the need for debriefing experiment participants.

\subsection{Direct Harm}

The possibility of direct harm to participants also forms an ethical
challenge.  Given the breadth of behavior that bots can exhibit, there
are many ways that bots could cause harm, and generating an exhaustive
list is impractical.  A few potential sources that are likely to commonly arise are
exposure to explicit or disturbing content, direct manipulation of
behavior in a negative way, violation of trust, and inconvenience or
annoyance.  Care must be taken in experimental design in order to
avoid these and other potential sources of direct harm.

\subsection{Terms of Service}

Terms of service may too pose a potential ethical question for
research involving bots.  Certain websites, such as Twitter,\footnote{https://support.twitter.com/articles/76915}
allow account automation for certain purposes, but other sites do not.
Facebook's terms of service forbids accessing the site through
automated means without prior
permission,\footnote{https://www.facebook.com/terms} and Yik Yak
forbids automated access
altogether.\footnote{https://www.yikyak.com/terms} The key question
in these cases is whether it is unethical to violate the site's terms
of service in order to conduct a scientific experiment.

There is significant controversy regarding this question
\cite{vitak2016beyond}.  The ACM Code of Ethics states that
``violations [of the terms of a license agreement] are contrary to
professional behavior'', and these violations may even be considered
illegal under the Computer Fraud and Abuse Act
\cite{vaccaro2015agree}.  At the same time, researchers have
challenged this guidance on the grounds that some types of important
research (e.g., in algorithmic accountability) are impossible without
violating terms of service \cite{vaccaro2015agree}, and in at least
one court case in the United States, violating terms of service was ruled as
non-criminal \cite{hofman2010court}.  


In light of these conflicting perspectives, we recommend following
terms of service unless the research question absolutely cannot be
answered without violating them, and only if the benefits of the study
outweigh the risks of this harm.  The intention to violate terms of service should also be approved by the researcher's Institutional Review Board (IRB).


\subsection{Context Dependence}

A final ethical challenge is context dependence.  Ethical
considerations always vary from case to case, and context is
especially important in reasoning about internet ethics
\cite{nissenbaum2004privacy}.  Unlike laboratory experiments, where
the context of an experiment is relatively easy to control, the
context of field experiments varies greatly from one to the next.  A
confedabot experiment on Twitter may be received quite differently
from an analogous experiment conducted on reddit due to the differing
norms on these sites, or the differing expectations that users have
about each other.

One important consideration in deciding whether to conduct an
experiment involving confedabots on a particular website is the
current presence or absence of bots on the site.  Some online
communities, such as Wikipedia and Twitter, have norms that are
relatively welcoming to the presence of bots.  In other communities,
such as Facebook or Yik Yak, the presence of a bot may be unusual. In
order to avoid violating the expectations or trust of a website's
users, and simply to avoid annoying them, it is important to observe
whether bots are already present on that site and to observe how those
bots typically participate in that space.  Contingency on context is
especially important in evaluating whether to use deception.  In an
environment where there are no bots, or where honesty and integrity in
the community is held as a standard for participation, deception may
be less justifiable.

Further complicating matters, many large websites also have a diverse
enough user base that there may be multiple communities with highly
disparate norms within the site.  For example, psychological responses
are known to vary widely between cultures \cite{henrich2010most}.
These cultural differences may play an important role in the ethics of
online experimentation, and should be understood better.  Anecdotal
evidence in prior work suggests that there may be cultural variation
in receptivity to bots \cite{savage2016botivist}.

Other contextual factors include the nature of the interactions
involved in the experiment at hand.  In a real example from early
internet research, a support group website's users grew upset when
they learned researchers were watching them
\cite{kraut2004psychological}.  In other cases, the research context
might put participants at a higher risk of harm, such as an experiment
mediating between abusive individuals and victims
\cite{matias2015reporting}.  

There is little controversy that risks should be carefully considered
in deploying any system, and context is an important part of that
process.  The ACM Code of Ethics charges us to ``give comprehensive
and thorough evaluations of computer systems and their impacts,
including analysis of possible risks''.  The AoIR states ``ethical
decision-making is best approached through the application of
practical judgment attentive to the specific context.''

To meet this challenge, we suggest carefully researching the norms of
user behavior on whatever site is being used to conduct a confedabot
experiment.  In the style of the AoIR's guidelines, some useful
questions to ask are: ``Do bots exists already on this site?'' ``Do
other users of the site know that bots exist?''  ``What are typical
behaviors of the users?''  ``What are typical behaviors of the bots?''
``Who are the users that are most likely to be interacting with the
bots involved in this research?''  ``Are those users more or less
likely to be upset that they are interacting with bots?''









\section{Proposed Guidelines}

We now propose a set of conservative guidelines for meeting the status
quo for ethical behavioral experimentation motivated by the issues we
have discussed.  Given that informed consent is likely to be difficult
to obtain in confedabot experiments, the main principle behind these
guidelines is achieving minimal risk as defined in the Common Rule.
In brief, the
guidelines we suggest are:
\begin{enumerate}
\item
  \textbf{Ordinary Behavior:} The use of a confedabot should not expose
  people to anything they would not be exposed to anyway.  In
  particular,
  \begin{enumerate}
  \item
    A confedabot should only be used in communities where bots are
    already present.
  \item
    The actions and attributes of a confedabot should not be unusual.
  \end{enumerate}
\item
  \textbf{Harm Reduction:} Efforts should be made to ensure a
  confedabot does not cause direct harm.
\item
  \textbf{Careful Evaluation:} The potential impact and potential harm
  from using confedabots should be carefully evaluated in the context
  of the site of the proposed research.
\end{enumerate}

\subsection{Ordinary Behavior}

The Common Rule defines minimal risk as meaning ``the probability and
magnitude of harm or discomfort anticipated in the research are not
greater in and of themselves than those ordinarily encountered in
daily life''.  To comply, bots should not alter the overall experiences of other
users on the site of the experiment.  If bots are not already present
on a site, then creating confedabots on that site risks exposing
people to bots for the first time in that environment.  Therefore
confebots are most appropriately used on sites that already have a noticeable
bot presence.  We also recommend avoiding unusual behavior, such
as inordinately high levels of activity or directly messaging random
strangers, since these behaviors could also alter users experiences.
We therefore propose as a guideline that confedabots should act like
``ordinary users'' on sites that already have bots.







\subsubsection{Example}

One example of a confedabot that conforms to this guideline is a bot
that upvotes random content on reddit.  To meet the guideline, this
experiment should only be conducted in subreddits where bots are used.
Since upvoting occurs frequently on reddit, this simple bot is not
performing any unusual actions, especially if the number of upvotes is
only a small number per hour.

\subsection{Harm Reduction}

We recommend implementing infrastructure that reduces the potential
direct harm a confedabot could cause.  Even if a bot's behavior is
ordinary on a particular site, many actions that might be normal on
that site could still be considered harmful.  For instance, verbal
abuse and harassment are common in certain online contexts
\cite{matias2015reporting}.  Methods such as persistently targeting
particular users should probably be avoided.  Mechanisms such as
keyword filters and curated content libraries should be put in place
to avoid posting disturbing or offensive content.  Explicit deception,
including misrepresenting facts or being unduly personal, is most safe
to avoid.  A form of harm reduction for interventions in especially
sensitive research areas is semi-automation.  In this case the bot
would have partial human supervision so that human judgement could
play a role in deciding when a particular intervention might be too
risky.

\subsubsection{Example}

One example of a failure to meet this guideline is Microsoft's Tay
bot.  Few details are available about how Tay was implemented, but
from the bot's behavior, it appears that minimal precautions were
taken to avoid posting disturbing content.  Much of the most extreme
behavior Tay exhibited could have likely been avoided using
keyword-based content filters.

An example of semi-automation for harm reduction is a recent
experiment in which researchers varied race and status attributes of
Twitter accounts and observed how these characteristics affected
response to censure for the offensive use of racial slurs
\cite{mungertweetment}.  The experiment arguably did not use bots
since the Twitter accounts were not autonomously controlled, but the
experiment did involve automation for detecting users to target in the
experiment, and also integrated human supervision of the subject
population, for example to help ensure that minors were not targeted.


\subsection{Careful Evaluation}

Our final recommendation is that the costs and benefits of using
confedabots should be carefully evaluated.  The particulars of the
proposed experimental design and the context of the site of the
proposed experiment should be carefully considered in these
evaluations.  In the words of the AoIR \cite{markham2012ethical}:
\begin{quote}
Ethical decision-making is a deliberative process, and researchers
should consult as many people and resources as possible in this
process, including fellow researchers, people participating in or
familiar with contexts/sites being studied, research review boards,
ethics guidelines, published scholarship (within one’s discipline but
also in other disciplines), and, where applicable, legal precedent.
\end{quote}
As a component of these considerations, if it is possible to perform
the experiment in a way that allows for informed consent or a more
laboratory-like online setting, then that route should be explored,
and existing guidelines (e.g., \cite{kraut2004psychological}) should
be employed.  For example, if a website is designed specifically for
the purposes of the experiment, or if the researcher has a
collaboration with the owners of the website that is used for the
experiment at hand, blanket opt-in decisions may be possible for users
of the website.

\subsubsection{Example}

One exemplar of all the above guidelines is the recent work on
``botivists'' that explored the best social media strategies to use
for recruiting volunteer activists \cite{savage2016botivist}.  The
experiment was conducted on Twitter, where bots are already common,
and the actions of the bot were not unusual (e.g., they did not make
an inordinate number of posts or disrupt existing discussions or make
outrageous requests).  The authors also took great care to avoid harm,
such as in refraining from following up with users who never
interacted with their bots.  Finally, the authors were thoughtful in
their assessment of the harms and benefits of their experiment.  For
example, the authors were cognizant of the unanticipated distress that
the bots caused to some of the activist community members.

\section{Discussion and Conclusions}

Our main goals in this paper were to articulate the methodology of
bots as virtual confederates for online field experiments, to outline
a design space for ``confedabots'', and to anticipate and
preemptively address the ethical issues that arise with conducting this type
of experiment.  Our hope is that this discussion will encourage and advance
experimental work using this methodology.

Significant gaps remain in this initial exploration of the ethics of
confedabots. Our guidelines were targeted at meeting the current
conditions for ethical experimentation. This approach is likely enough
for approval from an IRB or its equivalent but does not provide us
with a way to meet the strictest criteria outlined by the Belmont
Report. In particular current standards are heavily weighted towards a
utilitarian comparison of risk and harm, where a small probability of
harm is justifiable and minor violations of personal autonomy can be
permissible. The extent to which these transgressions cause actual
harm is an open question. For instance it is unknown the extent to
which online field experiments conducted without informed consent
might surprise or offend online users. Even if the experiment has no
potential harm to participants, there is still the possibility that
people will react negatively to the use of experimentation in their
online community.  The Facebook emotional contagion experiment
controversy suggests that some users might react negatively to any
experiment, no matter how many precautions have been taken.  At the
same time, an aspect of that backlash might have been that Facebook
was directly involved in the experiment.  Perhaps confedabots would be
treated as more permissible in cases when they are implemented by
peers on the system.

Another question is the extent to which the public at large
understands the pervasiveness of bots in online spaces. Our guidelines
suggest limiting confedabots to sites where bots are pervasive
already. A stricter guideline would only allow bots where people are
actually aware of the presence of those bots. How to most effectively
evaluate the relevant norms on a website in order to minimize harm is
also unclear. Shifts in public perceptions on these issues may also
occur over time, and tracking these changes presents a further
challenge.

Ethical concerns regarding the use of confedabots point to a broader need for future efforts to elaborate ethical guidelines for online experimentation. A possible strategy would be to create a system for maintaining an open public dialogue about social science and the importance of experimentation. Such a system could mitigate the most serious potential risks of omitting informed consent and using deception. On the scientific and methodological side, future work could instantiate or expand the design space we have proposed.

\section{Acknowledgments}

This work was supported in part by the NSF GRFP under grant
\#1122374, by NSF grants SES-1303533 and SES-1226483, and by the National Research Foundation of Korea grant NRF-2013S1A3A2055285. Any opinions, findings, and conclusions or recommendations
expressed in this material are those of the authors and do not
necessarily reflect those of the sponsors.


%
%
%
%
%
\balance{}

\bibliographystyle{SIGCHI-Reference-Format}
\bibliography{bot-ethics}

\end{document}